\documentclass[journal]{IEEEtran}

\usepackage{amsmath}
\allowdisplaybreaks[4]
\usepackage{amssymb}
\usepackage{amsthm}
\usepackage{cases}
\usepackage{bm}
\usepackage{algorithm}
\usepackage{algorithmic}
\usepackage{cases}
\usepackage{subfigure}
\usepackage{multirow}
\usepackage{xcolor} 
\usepackage{enumerate}
\usepackage{etoolbox}
\usepackage{boxedminipage} 
\usepackage{booktabs} 
\usepackage{graphicx}
\usepackage{mathrsfs}
\usepackage{caption}
\usepackage[pagebackref=true,breaklinks=true,letterpaper=true,colorlinks,bookmarks=false]{hyperref}

\newcommand{\MyMapTemplatePrefix}[4]{\expandafter#1\csname#3#4\endcsname{#2{#4}}}
\newcommand{\MyMapTemplatePrefixNew}[5]{\expandafter#1\csname#4#5\endcsname{#2{#3{#5}}}}
\forcsvlist{\MyMapTemplatePrefix {\def} {\mathbf} {}} {A,B,C,D,E,F,G,H,I,J,K,L,M,N,O,P,Q,R,S,T,U,V,W,X,Y,Z} 
\forcsvlist{\MyMapTemplatePrefix {\def} {\mathbf} {}} {a,b,c,d,e,f,g,h,j,k,l,m,n,o,p,q,r,s,t,u,v,w,x,y,z,1,0}
\forcsvlist{\MyMapTemplatePrefix {\def} {\widetilde} {wt}} {A,B,C,D,E,F,G,H,I,J,K,L,M,N,O,P,Q,R,S,T,U,V,W,X,Y,Z}
\forcsvlist{\MyMapTemplatePrefix {\def} {\widetilde} {wt}} {a,b,c,d,e,f,g,h,i,j,k,l,m,n,o,p,q,r,s,t,u,v,w,x,y,z} 
\forcsvlist{\MyMapTemplatePrefixNew {\def} {\widetilde}{\mathbf} {tb}} {A,B,C,D,E,F,G,H,I,J,K,L,M,N,O,P,Q,R,S,T,U,V,W,X,Y,Z}
\forcsvlist{\MyMapTemplatePrefixNew {\def} {\widetilde}{\mathbf} {tb}} {a,b,c,d,e,f,g,h,i,j,k,l,m,n,o,p,q,r,s,t,u,v,w,x,y,z}
\forcsvlist{\MyMapTemplatePrefix {\def} {\widehat} {wh}} {A,B,C,D,E,F,G,H,I,J,K,L,M,N,O,P,Q,R,S,T,U,V,W,X,Y,Z}
\forcsvlist{\MyMapTemplatePrefix {\def} {\widehat} {wh}} {a,b,c,d,e,f,g,h,i,j,k,l,m,n,o,p,q,r,s,t,u,v,w,x,y,z}
\forcsvlist{\MyMapTemplatePrefixNew {\def} {\widehat}{\mathbf} {hb}} {A,B,C,D,E,F,G,H,I,J,K,L,M,N,O,P,Q,R,S,T,U,V,W,X,Y,Z}
\forcsvlist{\MyMapTemplatePrefixNew {\def} {\widehat}{\mathbf} {hb}} {a,b,c,d,e,f,g,h,i,j,k,l,m,n,o,p,q,r,s,t,u,v,w,x,y,z}
\forcsvlist{\MyMapTemplatePrefix {\def} {\mathcal}{mc}} {A,B,C,D,E,F,G,H,I,J,K,L,M,N,O,P,Q,R,S,T,U,V,W,X,Y,Z,a,b,c,d,e,f,g,h,i,j,k,l,m,n,o,p,q,r,s,t,u,v,w,x,y,z}
\forcsvlist{\MyMapTemplatePrefix {\def} {\mathbb} {mb}} {A,B,C,D,E,F,G,H,I,J,K,L,M,N,O,P,Q,R,S,T,U,V,W,X,Y,Z}
\forcsvlist{\MyMapTemplatePrefix {\DeclareMathOperator} {} {} } {tr,diag,sgn}

\begin{document}

\title{Multi-level Attention-guided Graph Neural Network for Image Restoration}

\author{Jiatao~Jiang,
        Zhen~Cui,
        Chunyan~Xu
        and~Jian Yang}

\markboth{IEEE TRANSACTIONS ON NEURAL NETWORKS AND LEARNING SYSTEMS, February~2025}%
{Shell \MakeLowercase{\textit{et al.}}: Bare Demo of IEEEtran.cls for IEEE Journals}

\maketitle

\begin{abstract}
In recent years, deep learning has achieved remarkable success in the field of image restoration. However, most convolutional neural network-based methods typically focus on a single scale, neglecting the incorporation of multi-scale information. In image restoration tasks, local features of an image are often insufficient, necessitating the integration of global features to complement them. Although recent neural network algorithms have made significant strides in feature extraction, many models do not explicitly model global features or consider the relationship between global and local features. This paper proposes multi-level attention-guided graph neural network. The proposed network explicitly constructs element block graphs and element graphs within feature maps using multi-attention mechanisms to extract both local structural features and global representation information of the image. Since the network struggles to effectively extract global information during image degradation, the structural information of local feature blocks can be used to correct and supplement the global information. Similarly, when element block information in the feature map is missing, it can be refined using global element representation information. The graph within the network learns real-time dynamic connections through the multi-attention mechanism, and information is propagated and aggregated via graph convolution algorithms. By combining local element block information and global element representation information from the feature map, the algorithm can more effectively restore missing information in the image. Experimental results on several classic image restoration tasks demonstrate the effectiveness of the proposed method, achieving state-of-the-art performance.
\end{abstract}


\IEEEpeerreviewmaketitle

\section{Introduction}
In real-world scenarios, digital images often suffer from information loss due to environmental factors, equipment limitations, or human errors. For instance, motion blur caused by moving objects, noise introduced by damaged cameras, resolution degradation from image compression algorithms, and occlusions resulting from manual modifications are common issues. Restoring low-quality images to high-quality ones by mitigating these factors is a fundamental task in image processing. Image restoration (IR) aims to reconstruct high-quality images by eliminating degradations such as noise, blur, occlusions, missing data, and raindrops. Due to the inherently uncertain nature of image restoration, reconstructing high-quality images remains a highly challenging task.

In the study and application of image restoration, tasks can be categorized into several classic sub-tasks: image denoising~\cite{buades2005review}, image reconstruction~\cite{gull1978image}, compression artifact removal~\cite{dong2015compression}, demosaicing~\cite{kimmel1999demosaicing}, super-resolution~\cite{dong2015image}, and compressed sensing~\cite{lustig2008compressed}. The image restoration process can be expressed as:
\begin{equation}
	\mathcal{I}_H = \mathcal{I}_L + n,
\end{equation}
where $\mathcal{I}_H$ and $\mathcal{I}_L$ represent high-quality and low-quality images, respectively, and $n$ denotes the complementary information required to reconstruct the high-quality image from the low-quality one. In practice, the observed image is typically $\mathcal{I}_L$, and the goal is to learn the complementary information $n$ through model algorithms to reconstruct the high-quality image.

Generally, image restoration methods can be divided into two categories: traditional model-based approaches and popular neural network-based approaches. Traditional methods primarily rely on spatial filters~\cite{tomasi1998bilateral}, domain transformation techniques~\cite{starck2002curvelet}, and sparse representation-based methods~\cite{buades2005review,buades2005non,dong2011sparsity}. These methods model specific characteristics of the original image, analyzing and identifying patterns to reconstruct high-quality images. For example, the classic BM3D algorithm~\cite{dabov2007image} (Block-matching and 3D Filtering) enhances sparse representation in the transform domain by grouping similar image patches and applying 3D filtering. Additionally, methods based on sparse representation and low-rank constraints~\cite{dong2012nonlocally,gu2014weighted} have also achieved significant success. However, traditional methods have notable limitations: i) they often require iterative solutions, leading to high computational costs; ii) they may involve manually tuned parameters; and iii) they are typically tailored to specific restoration tasks and lack versatility.

In contrast, deep learning-based models, particularly convolutional neural networks(CNNs)~\cite{zhang2017beyond,zhang2018ffdnet,plotz2018neural,liu2018non,valsesia2020deep,valsesia2020deep,mou2021dynamic}, have achieved remarkable success in image restoration due to their powerful representation capabilities. These methods model the mapping relationship between low-quality and high-quality images, leveraging the strong learning ability of neural networks to infer the complementary information $n$. For instance, the Trainable Nonlinear Reaction Diffusion (TNRD) method~\cite{chen2016trainable} uses multi-layer perceptron networks for image denoising, achieving excellent performance through extensive training. Similarly, the Denoising Convolutional Neural Network (DnCNN)~\cite{zhang2017beyond} employs batch normalization~\cite{ioffe2015batch} and residual connections to enhance denoising performance. CNNs excel in feature extraction and generalization, using convolutional operations to extract image representations and stacking multiple layers to capture high-dimensional semantic information. Additionally, the self-similarity property of images is utilized in network construction and learning. By propagating non-local information, local content in low-quality images can be restored using other parts of the image, enabling the model to better learn missing information. For example, the Neural Nearest Neighbors Network (N3Net)~\cite{plotz2018neural} leverages nearest-neighbor matching of patch features to fuse non-local information, outperforming standard CNNs and similar non-local methods.

Furthermore, the rise of graph convolutional networks (GCNs) has expanded the range of methods applicable to image restoration. Valsesia et al. proposed a Graph Convolutional Denoiser Network (GCDN)~\cite{valsesia2020deep}, which dynamically computes graph adjacency relationships based on feature similarity and employs a lightweight edge convolutional network to avoid gradient explosion or vanishing and over-parameterization. Mou et al. introduced a Dynamic Attentive Graph Learning (DAGL) framework~\cite{mou2021dynamic}, which transmits and aggregates local structural information through a graph network embedded in a residual network for end-to-end learning. However, traditional CNNs often focus on information within the receptive field, neglecting the analysis and integration of global feature map information. While non-local neural network methods fuse global information, they typically involve large computational costs or parameters and lack a unified design framework.

\begin{figure*}[t]
	\centering
	\includegraphics[width=0.9\textwidth]{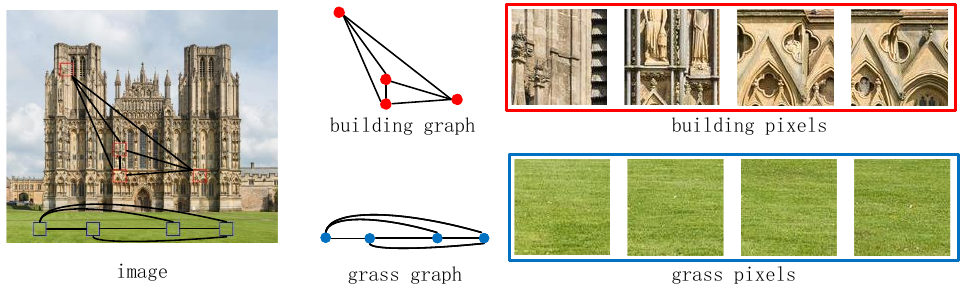}
	\caption{The diagram illustrates the graph structure constructed between pixel patches in an image. The four red boxes highlight local patches of a building, which exhibit highly similar representation and structural information. Similarly, the four blue boxes highlight local patches of grass, which also share analogous features. For both the building and grass patches, a connected graph can be constructed to propagate and aggregate local structural information and global representation information across the image. This process supplements and refines the missing or incomplete details in each patch by leveraging the shared characteristics of similar regions. Through this graph-based approach, the network effectively integrates local and global information, enhancing the overall restoration and representation of the image.}
	\label{fig:motivation}
\end{figure*}

To address the limitations of GCNs in extracting local features and integrating global representations, this chapter proposes a Multi-level Attention-guided Graph Network (MAGN). Inspired by DAGL~\cite{mou2021dynamic}, MAGN is a multi-level GCN comprising a global representation graph and a local structure graph. As illustrated in Figure~\ref{fig:motivation}, local patches of buildings (red boxes) and grass (blue boxes) exhibit highly similar structural and pixel representation features. For low-quality images, similar global representations and structures can be leveraged to restore missing local information. MAGN constructs connected graphs for building and grass patches, propagating and aggregating local structural and global representation information. The global representation graph enables each feature to reference and update based on overall information, while the local structure graph allows small regions to reference and update based on all local structures. Additionally, a multi-head attention mechanism dynamically constructs representation and structure graphs, enhancing the network's expressive power.

In summary, this chapter proposes an attention-guided graph convolutional network for image restoration. The main contributions are as follows:
\begin{itemize}
	\item Introducing GCNs to image restoration tasks, enabling better learning and utilization of non-local image information.
	\item Proposing a multi-level graph structure that flexibly combines image representation and structural information, using a multi-head attention mechanism to dynamically learn graph adjacency relationships and fuse non-local information, thereby improving feature extraction performance.
	\item Demonstrating the effectiveness of the proposed method through quantitative experiments on demosaicing, synthetic image denoising, and compression artifact removal tasks, with results across multiple datasets highlighting its robustness and state-of-the-art performance.
\end{itemize}

\section{Related Works}
Recently, due to the significant advantages of deep networks in the field of image restoration, convolutional neural network (CNN)-based methods have garnered increasing attention and research. Dong et al. proposed a deep neural network for artifact reduction~\cite{dong2015compression} (Artifacts Reduction Convolutional Neural Networks, ARCNN). ARCNN extracts effective image representations using a deep convolutional neural network and learns a network-based mapping structure for image restoration. Zhang et al. introduced a fast and flexible denoising convolutional neural network~\cite{zhang2018ffdnet} (Fast and Flexible Denoising Convolutional Neural Network, FFDNet). FFDNet employs a multi-layer convolutional network with batch normalization to extract and fuse features from downsampled input images and noise level maps, accelerating training and enhancing the network's expressive power. Zhang et al. proposed a CNN-based image restoration method~\cite{zhang2017learning} (Image Restoration CNN, IRCNN). IRCNN analyzes the strengths and weaknesses of model-based and learning-based approaches, combining their advantages to improve network performance in image restoration tasks. Liu et al. developed a novel multi-level wavelet CNN~\cite{liu2018multi} (Multi-level Wavelet CNN, MWCNN). MWCNN integrates wavelet decomposition with CNNs, designing a multi-level wavelet convolutional network and incorporating it into a U-shaped CNN architecture. Tian et al. proposed an attention-guided CNN~\cite{tian2020attention} (Attention-guided CNN, ADNet). ADNet uses an attention mechanism to more accurately extract noise information and employs dilated convolution~\cite{yu2015multi} to balance network performance and efficiency. Anwar et al. introduced a feature attention-based method~\cite{anwar2019real} (Real Image Denoising Network, RIDNet). RIDNet constructs long and short skip connections to learn global and local feature map residuals, respectively. The residual design in RIDNet effectively prevents gradient explosion or vanishing, ensuring stable training, while the skip connections enable complementary information exchange between low-level representations and high-level semantics. Additionally, Chupraphawan et al. proposed a method leveraging edge images~\cite{chupraphawan2019deep} (Deep Convolutional Neural Network with Edge Feature, DCEF). DCEF addresses the long convergence time in deep network training by introducing an adaptive learning rate based on triangular techniques, enabling faster convergence. It also incorporates edge image information into the denoising task, improving edge quality in restored images. Fang et al. proposed a multilevel edge features-guided network~\cite{fang2020multilevel} (Multilevel Edge Features Guided Network, MLEFGN). MLEFGN integrates edge detection, edge guidance, and image denoising into a single end-to-end network, predicting edge information from noisy images and using it as a prior for low-quality images. This approach enhances the network's accuracy and robustness. Liu et al. introduced a gradient-based method~\cite{liu2020gradnet} (Gradient Network, GradNet). GradNet fuses gradient information computed from a prior network with shallow image features, preserving high-frequency textures and edges to improve restoration quality. It also proposes a gradient consistency regularization to enforce gradient similarity between denoised and high-quality images, ensuring that the restored images are closer to the ground truth.

Recently, the rise of graph convolutional networks (GCNs) has expanded the range of methods applicable to image restoration tasks. Valsesia et al. proposed a graph convolutional denoiser network~\cite{valsesia2020deep} (Graph Convolutional Denoiser Network, GCDN). GCDN dynamically computes graph adjacency relationships based on feature similarity and employs a lightweight edge convolutional network to avoid gradient explosion or vanishing and over-parameterization, ensuring stable and efficient learning. Mou et al. introduced a dynamic attentive graph learning framework~\cite{mou2021dynamic} (Dynamic Attentive Graph Learning, DAGL). DAGL transmits and aggregates local structural information through a graph network embedded in a residual network for end-to-end learning. The graph structure in DAGL supplements missing information in low-quality images by leveraging local structural information from the entire feature map, improving network accuracy. However, traditional CNNs often focus on information within the receptive field, neglecting the analysis and integration of global feature map information. While non-local neural network methods fuse global information, they typically involve large computational costs or parameters and lack a unified design framework, requiring manual intervention.

\section{Method}
We proposed a graph convolutional neural network with multi-headed attention to learn better representation information than traditional convolutional neural network.
The framework is an end-to-end network including residual block module and graph convolution module.
The graph convolution module contain two submodules: i) Pixel level information aggregation module; ii) Patch level information aggregation module.
The more details can be found in next section.

\subsection{Framework}
\begin{figure*}[t]
	\centering
	\includegraphics[width=1.0\textwidth]{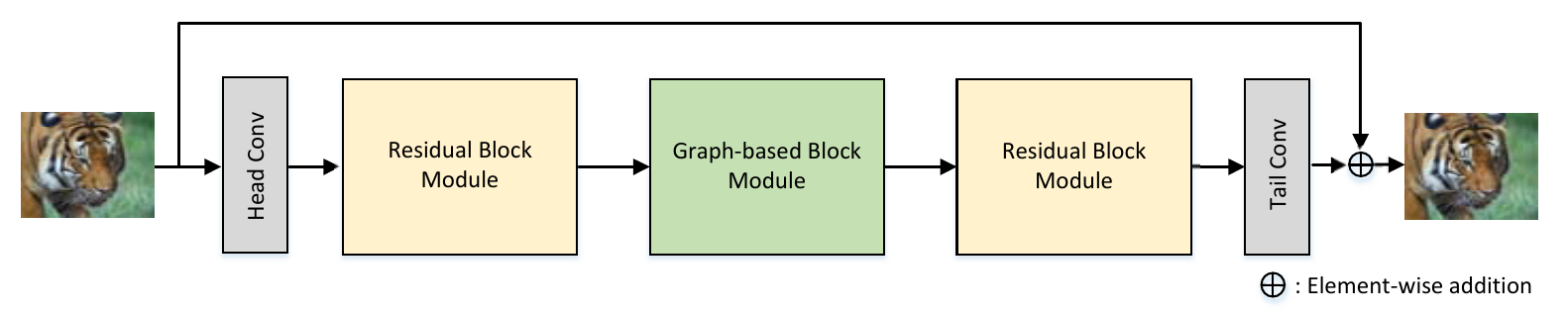}
	\caption{The framework of our proposed model.
		The network contains two module: i) residual block module; ii) graph-based block module.
		The residual block module stacks some residual blocks to learn the residual information.
		The graph-based block module mainly employs graph convolution with multi-headed attention to learn the interation information between the pixel and patch on images.
		The framework learns a residual of the original image to restore the high quality image.}
	\label{fig:framework}
\end{figure*}

\begin{figure}[t]
	\centering
	\subfigure[Residul block module]{
		\label{fig:resblock}
		\includegraphics[width=0.36\textwidth]{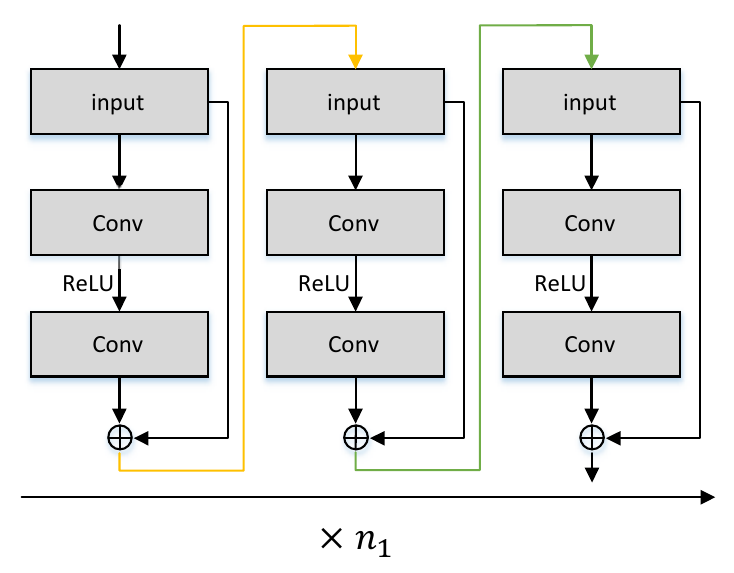}
	}
	\subfigure[Graph-based block module]{
		\label{fig:graphblock}
		\includegraphics[width=0.405\textwidth]{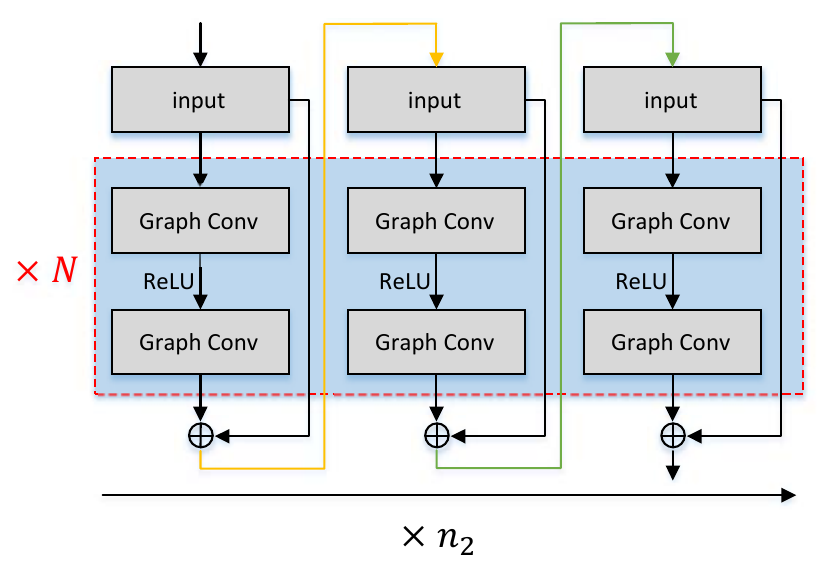}
	}
	\caption{The comparison of the two basic modules.
		The residual module employs two convolution layers to learn residuals of the input and stacks $n_1$ basic residual blocks.
		The graph-based module employs two graph convolution layers to learn residuals of the input and has $n_2$ graph-based blocks.
		The graph-based block propagates interactive information on images with $N$ multi-headed attention.}
	\label{fig:block}
\end{figure}

\begin{figure}[t]
	\centering
	\includegraphics[width=0.5\textwidth]{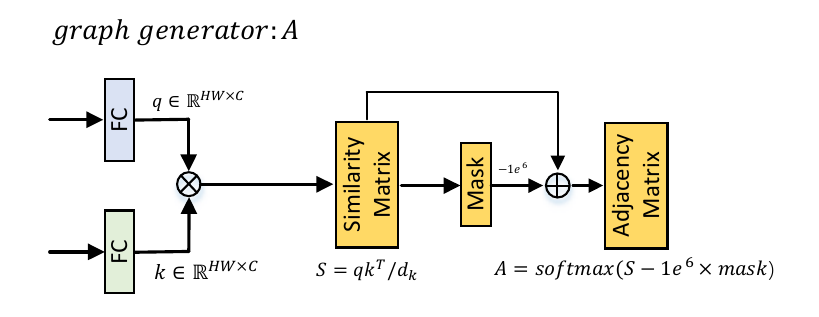}
	\caption{Graph generator.
		The generator mainly employ the attention mechanism to construct a adjacency matrix of inputs.
		The feature map is projected to representation space $q, k$ by the representation function.
		Then, similarity matrix $\mathbf{S}$ is calculated by the representation feature $q, k$.
		What's more, the mask matrix avoid the disturbance of low similarity nodes.
		Finally, the adjacency matrix $\mathbf{A}$ contains the interaction relations between the input data.}
	\label{fig:adj}
\end{figure}

An overview of our proposed model is shown in Fig.~\ref{fig:framework}.
The framework is an end-to-end residual network including two main modules: i) residual block module; ii) graph-based block module.
Firstly, the network transforms the input image into feature space by the head convolution layers.
Secondly, features are projected into the residual space of low quality images by stacked residual block module and graph-based block module, iteratively.
Finally, the tail convolution layer reduces dimension of the residual space to obtain the real residual value.
Then, the residual network computes the estimation for the high quality images.
Given the low quality images $\mathcal{I}_{L}\in \mathbb{R}^{W\times H\times C}$, supposed the residuals can be learnt by the proposed model $\mathcal{F}$,
we have the reconstruction $\mathcal{I}_{H}\in \mathbb{R}^{W\times H\times C}$ of the high quality images can be denoted as following:
\begin{align}
	\mathcal{I}_H = \mathcal{I}_L + \mathcal{F}(\mathcal{I}_L).
\end{align}

\subsection{Graph-based block module}
As shown in Fig.~\ref{fig:block}, the residual block module and graph-based block module stack many residual blocks in the network, iteratively.
Given the inputs $x\in \mathbb{R}^{W\times H\times C}$, residual block and convolution layers $\text{ResBlock}, f_1, f_2$,
graph block and graph convolution layers $\text{GraphBlock}, g_1, g_2$, activation function $\sigma$,
outputs of basic residual block and graph-based block are denoted as following:
\begin{align}
	x_1 &= \text{ResBlock}(x) = f_2(\sigma(f_1(x))) + x, \nonumber \\
	x_1 &= \text{GraphBlock}(x) = g_2(\sigma(g_1(x))) + x.
\end{align}
As shown in Fig.~\ref{fig:resblock}, the residual block learns the residuals of the inputs by two convolution layers $f_1, f_2$.
Meanwhile, the module can better extract the image information by stacked $n_1$ residual blocks.
However, the graph-based block employs two graph convolution layers $g_1, g_2$ to propagate information.
What's more, the graph convolution layer learns the adjacent relations between pixels and patches on images by $N$ multi-headed attention mechanisms, shown in Fig.~\ref{fig:graphblock}.
Next, the residual connection is applied after the two graph convolution layers.
Further, the graph-based module also stacks $n_2$ block to enrich the features.

\subsubsection{Graph convolution with multi-headed attention} \label{sec:graphconvolution}

Given the directed/undirected graph $\mathcal{G} = (\mathcal{V}, \mathbf{A}, \mathbf{X})$.
$\mathcal{V} = \{v_i\}_{i=1}^{m}$ is a finite set of $m$ nodes on graph,
$\mathbf{A} \in \mathbb{R}^{m \times m}$ is a adjacency matrix to record the relations between nodes and others,
$\mathbf{X} \in \mathbb{R}^{m \times d}$ is a attribute matrix with $d$ dimension feature for every node on graph.
We can employ the popular method GCN~\cite{kipf2016semi} or GAT~\cite{liu2017global} to aggregate the local information of the graph $\mathcal{G}$.
However, the output of the residual network is a feature map $\mathbf{X}$ and is an input for the graph block.
Thus, we proposed an approach to learn a dynamic adjacency matrix $\mathbf{A}$ by the attention mechanism, shown in Fig.~\ref{fig:adj}.
The input $\mathbf{X} \in \mathbb{R}^{m \times d}$ is regarded as the attribute matrix of $d$ dimension features in the graph block.

Supposed that we denote the inputs and softmax function as $q, k, v \in \mathbb{R}^{m \times d_1}$ and $\text{softmax}$ in the attention mechanism,
we have the output $v_{out}$ of attention as following:
\begin{align}
	v_{out} = \text{softmax}(qk^T/\sqrt{d_1})v.
\end{align}
In our proposed graph block, the attention mechanism is applied to construct the adjacency matrix.
The input $\mathbf{X}$ is projected into a feature space by an convolution layer in the graph block.
Then, we reduce the dimension of the feature space by the fully connected layer.
Given the convolution layer and fully connected layer $\text{Conv1}, \text{Conv2}, \text{FC1}, \text{FC2}$, we have the similarity matrix $\mathbf{S}$ as following:
\begin{align}
	q &= \text{FC1}(\text{Conv1}(\mathbf{X})), \nonumber \\
	k &= \text{FC2}(\text{Conv2}(\mathbf{X})), \nonumber \\
	\mathbf{S} &= qk^T/\sqrt{d}.
\end{align}
In fact, we calculate a mask for the similarity matrix, which reduce the computation complexity and avoid the disturbance of low similarity nodes.
The mask only is related to the similarity matrix and don't have parameters in the network, denoted as following:
\begin{align}
	\mathbf{mask} = \left\{
	\begin{array}{l l}
		1, &\text{if } \mathbf{S} < \overline{\mathbf{S}}, \\
		0, &\text{others}.
	\end{array}
	\right.
\end{align}
Here, $\overline{\mathbf{S}} \in \mathbb{R}^{m \times 1}$ is the mean value matrix of the similarity matrix.
Thus, we have the adjacency matrix $\mathbf{A} \in \mathbb{R}^{m \times m}$, denoted as following:
\begin{align}
	\mathbf{A} = \text{softmax}(\mathbf{S} - 1e^6\times\mathbf{mask}).
\end{align}
Here, $\mathbf{A}$ is a weighted and normalized adjacency matrix.
Note that, the GCN~\cite{kipf2016semi} employs the normalized adjacency matrix by the Laplacian matrix and normalization trick.
Hence, the input $\mathbf{X}$ is transformed into a graph $\mathcal{G} = (\mathbf{X}, \mathbf{A})$ by the proposed graph generator.
Accordingly, the output $\mathbf{X}_1$ of the graph convolution operation is denoted as following:
\begin{align} \mathbf{X}_1 = \sigma(g(\mathbf{X}, \mathbf{A})) = \sigma(\mathbf{A} \mathbf{X} \mathbf{W}).
\end{align}
Here, $\mathbf{W} \in \mathbb{R}^{d \times d}$ is a weight matrix, which are learnable filter parameters in the network.
$g$ is the graph convolution operator.

Moreover, the multi-headed attention mechanism is also employed for the graph generator.
Supposed that $N$ multi-headed attentions mechanism produce $N$ similarity matrices $\mathbf{S}_1, \mathbf{S}_2, \cdots, \mathbf{S}_N$,
we can obtain $N$ adjacency matrices and a output $\mathbf{X}_1$ of graph convolution:
\begin{align}
	\mathbf{S}_i &= q_i k_i^T/\sqrt{d_i}, \nonumber \\
	\mathbf{A}_i &= \text{softmax}(\mathbf{S}_i - 1e^6\times \mathbf{mask}_i), \nonumber \\
	\mathbf{X}_1 &= \sigma(g_1(\mathbf{X}, \mathbf{A}_1), g_2(\mathbf{X}, \mathbf{A}_2), \cdots, g_N(\mathbf{X}, \mathbf{A}_N),) \nonumber \\
	&= \sigma(\text{concat}(\mathbf{A}_1 \mathbf{X} \mathbf{W}_1, \mathbf{A}_2 \mathbf{X} \mathbf{W}_2, \cdots, \mathbf{A}_N \mathbf{X} \mathbf{W}_N)).
\end{align}
Here, $\{\mathbf{W}_i \in \mathbb{R}^{d \times (d/N)}\}_{i=1}^{N}$ are weight matrices and $\text{concat}$ is an operator to concatenate the sequence of features.
Therefore, the multi-headed attention mechanism can produce multiple adjacency matrices and parameters can be learnt by the gradient descent method in the mechanism.
That is to say, the proposed graph convolution with the multi-headed attention facilitates the construction of network architectures and the optimization of parameters in the network.

\subsubsection{Information aggregation on pixels and patches}

\begin{figure*}[t]
	\centering
	\includegraphics[width=0.8\textwidth]{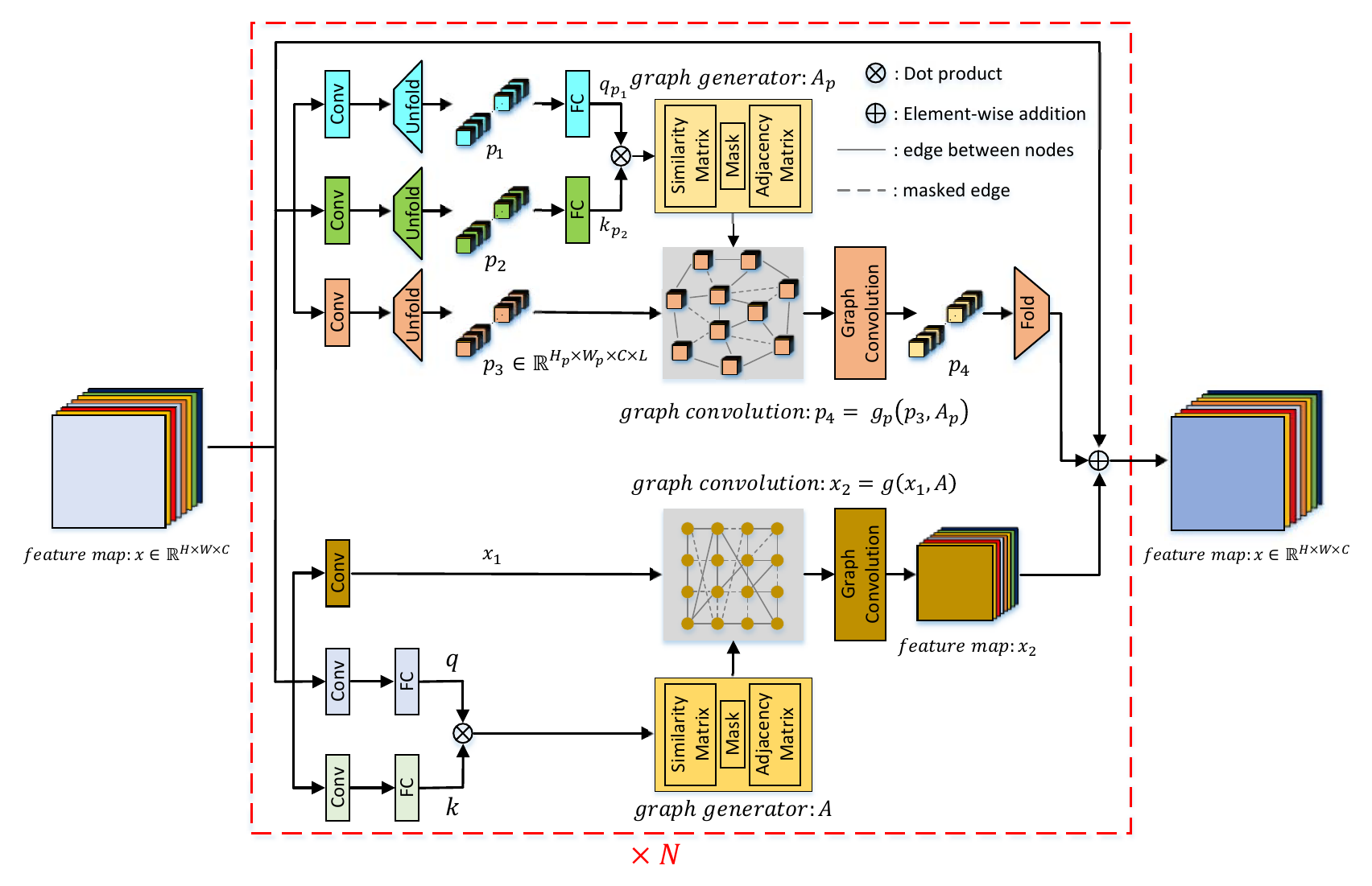}
	\caption{Information aggregation on pixels and patches.
		The bottom half is information aggregation on pixels.
		The graph generator constructs an adjacency matrix $\mathbf{A}$ with the attention mechanism on pixels.
		More details can be found in section~\ref{sec:graphconvolution}.
		The graph convolution operation extracts the global information and updates the feature map $\mathbf{X}_1$ on the pixel graph $\mathcal{G}=(\mathbf{X}, \mathbf{A})$.
		The top half is information aggregation on patches.
		The feature map is unfolded into patches $p_1 \in \mathbb{R}^{H_p\times W_p\times C\times L}$.
		Then, the adjacent relations $\mathbf{A}_p$ are modeled between $L$ patches.
		More details can be found in section~\ref{sec:graphconvolution}.
		The $L$ patches are folded into a feature map after the graph convolution operation $g_p(p_3, \mathbf{A}_p)$.
		Finally, two parts of the feature map are joined together as the residuals of the input.
		The output is the sum of the original feature map and the residuals on pixels and patches.
	}
	\label{fig:graphconv}
\end{figure*}
\begin{figure*}[t]
	\centering
	\includegraphics[width=0.7\textwidth]{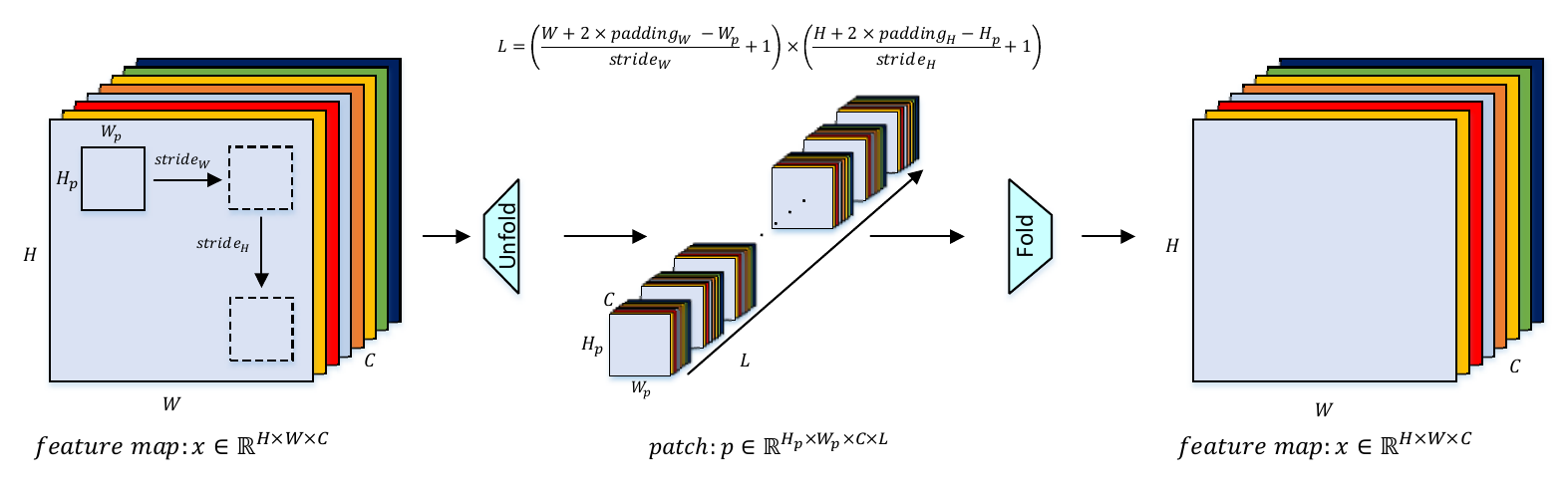}
	\caption{Patch generator.
		By sliding the window of $H_p \times W_p$ size and steps are $stride_H, stride_W$ in vertical and horizontal orientation,
		the feature map $\mathbf{X}\in \mathbb{R}^{H\times W\times C}$ is unfolded into $L$ patches.
		The size of the patch is $H_p\times W_p\times C$
		and the number $L = (\frac{H + 2\times padding_H - H_p}{stride_H} + 1)\times (\frac{W + 2\times padding_W - W_p}{stride_W} + 1)$.
		Here, $padding_H, padding_W$ are the number of paddings in vertical and horizontal orientation on images.
		Conversely, by the sliding the same window and steps, $L$ patches also are folded into a feature map $\mathbf{X}$.}
	\label{fig:fold}
\end{figure*}
The convolution operation can explicitly focus on the local information in the receptive field and extracts the effective features.
However, the convolution operation do not explicitly focus on global information.
The neural network implicitly obtains global information by expanding the receptive field and stacking multiple convolution layers.
To address this problem, graph convolution operation is employed on the information aggregation.
Due to explicit interaction relations, the graph convolution can focus on the global information in the feature map.
What's more, the graph convolution operation has few parameters and acceptable computational complexity on the network.

We proposed to model the global information on two receptive fields by the graph convolution operation, shown in Fig.~\ref{fig:graphconv}.
On the one hand, the minimal receptive field is a pixel on the image.
The image $\mathbf{I} \in \mathbb{R}^{H\times W\times 3}$ has $H\times W$ pixel.
We model the $H\times W$ spatial locations corresponding to pixels by the graph convolution with multi-headed attention.
The feature map $\mathbf{X}$ in the network is able to be transformed into a graph $\mathcal{G}=(\mathbf{X}, \mathbf{A})$ by the attention mechanism.
Then, the graph $\mathcal{G}$ is projected to a new feature map $\mathbf{X}_1$ by the graph convolution.
The new feature map $\mathbf{X}_1$ contains the global information and has same dimension with the input $\mathbf{X}$

On the other hand, the bigger receptive field is a patch on the image.
By sliding the window in the vertical and horizontal orientation, the image is unfolded into some patches.
The patch is exactly the size of a receptive field.
As shown in Fig.~\ref{fig:fold}, when the size of sliding window is $H_p\times W_p$ and the steps are $stride_H, stride_W$,
the image is segmented into $L$ patches:
\begin{align} \label{eqn:L}
	L = (\frac{H + 2\times padding_H - H_p}{stride_H} + 1)\times (\frac{W + 2\times padding_W - W_p}{stride_W} + 1).
\end{align}
Here, $H_p, W_p$ denotes the size of the patch, $padding_H, padding_W$ denotes the number of the padding on vertical and horizontal orientation.
Note that, if the $stride_H$/$stride_W$ is less than $H_p$/$W_p$, the patches overlap each other.
Then, the feature map $\mathbf{X}$ is able to be unfolded into $L$ patches by the same sliding window and steps.
As shown in Fig.~\ref{fig:graphconv}, the three feature maps are unfolded into three patches $p_1, p_2, p_3 \in \mathbb{R}^{H_p\times W_p\times C\times L}$.
Then, the graph generator calculates the similarity matrix, mask and adjacency matrix $\mathbf{A}_p$ by reducing the dimension of patches $p_1, p_2$.
The patch $p_3$ is transformed into a new feature $p_4$ by the graph convolution operation $g_p(p_3, \mathbf{A}_p)$.
Next, the new patch $p_4$ is folded into a feature map on the spatial location by same sliding window and steps.
The details of the graph convolution can be found in section~\ref{sec:graphconvolution}.

Finally, the two parts of the feature map are joined together as the residuals of the input feature map.
The output is the sum of the original feature map and the residuals on pixels and patches.

\subsubsection{Residual connection and layer normalization}
In the graph convolutional neural network framework for both global representation and local structure graphs, two distinct levels of residual outputs are obtained. These residuals are then combined with the original input features to form a new feature representation, as expressed below:
\begin{align}
	\mathbf{X} = \mathbf{X} + f(\underbrace{\text{Fold}(g_p(p_3, \mathbf{A}_p))}_{patch} + \underbrace{g(\mathbf{X}, \mathbf{A})}_{pixel})
\end{align}
Here, \( \mathbf{X} \) represents the input features, \( g_p \) and \( g \) denote the graph convolution operations for local and global information, respectively, and \( \mathbf{A}_p \) and \( \mathbf{A} \) are the adjacency matrices for local and global graphs. The function \( \text{Fold} \) aggregates the local structural information, while \( f \) integrates the combined residuals into the original features.

To ensure stable training and facilitate parameter learning, batch normalization is applied to the output features of the graph convolution:
\begin{align}
	\mathbf{X}_{norm} = \frac{\mathbf{X} - \mathbf{E}(\mathbf{X)}}{\sqrt{\mathbf{Var}(\mathbf{X})}}.
\end{align}
Here, \( \mathbf{E} \) and \( \mathbf{Var} \) represent the mean and variance operations, respectively. This regularization technique enhances data stability during training and mitigates issues such as gradient explosion or vanishing.

\subsubsection{Loss Function}
The image restoration task is approached in a supervised manner during experiments. Let the input low-quality image be denoted as \( \mathcal{I}_{L} \in \mathbb{R}^{W \times H \times 3} \), and the corresponding high-quality ground truth image as \( \mathcal{I}_{real} \in \mathbb{R}^{W \times H \times 3} \). Assuming the residual map can be learned by the proposed convolutional network \( \mathcal{F} \), the reconstructed high-quality image \( \mathcal{I}_{H} \in \mathbb{R}^{W \times H \times 3} \) can be expressed as:
\begin{align}
	\mathcal{I}_{H} = \mathcal{I}_{L} + \mathcal{F}(\mathcal{I}_{L}).
\end{align}
The objective function \( \mathcal{L} \) is then defined as:
\begin{align}
	\mathcal{L} = \frac{1}{M} \sum_{i=1}^{M} \lVert \mathcal{I}_{real}^i - \mathcal{I}_H^i \rVert _{2}^{2},
\end{align}
where \( M \) represents the total number of training samples. The mean squared error (MSE) is one of the most commonly used objective functions in image restoration tasks. Depending on the specific application, multiple loss functions can be combined to construct a more robust objective function for improved restoration performance.

\section{Experiments} 
The effectiveness of the proposed Multi-level Attention-guided Graph Neural Network (MAGN) for image restoration is evaluated on several key tasks, including synthetic image denoising, compression artifact reduction, and demosaicing. This section provides a detailed description of the experimental setup, results, and ablation studies.

\subsection{Network settings and datasets}
For the synthetic image denoising, compression artifact removal, and demosaicing tasks, the proposed MAGN model is trained on the DIV2K dataset~\cite{toivonen2003statistical}. DIV2K is a widely used dataset for single-image super-resolution, containing a large number of high-resolution color images with diverse content. The dataset consists of 1,000 images, divided into 800 training images, 100 validation images, and 100 test images. In this experiment, the 800 high-resolution training images are utilized.

\textbf{Synthetic Image Denoising Task}: Testing is conducted on the Urban100~\cite{huang2015single}, BSD68~\cite{martin2001database}, and Set12 datasets. Urban100 contains 100 images of urban scenes. BSD68 is a subset of natural images and specific objects, comprising 68 images. Set12 includes 12 grayscale images from various scenes.  

\textbf{Compression Artifact Removal Task}: Testing is performed on the Classic5~\cite{foi2007pointwise} and LIVE1~\cite{sheikh2005live} datasets. Classic5 consists of 5 grayscale images. LIVE1 is a dataset for image quality assessment, containing 29 images.  

\textbf{Demosaicing Task}: Testing is conducted on the McMaster18~\cite{zhang2017learning}, Kodak24, and Urban100~\cite{huang2015single} datasets. McMaster18 and Kodak24 are datasets specifically designed for demosaicing, containing 18 and 24 images, respectively. 

\begin{table*}[!h]
	\centering
	\small
	\caption[Detailed Configuration of Each Layer in the Attention-Guided Graph Convolutional Neural Network]{Detailed configuration of each layer in the Attention-Guided Graph Convolutional Neural Network ($W$, $H$, and $L$ (Equation~\ref{eqn:L}) represent the image width, height, and the number of graph nodes, respectively.}
	\begin{tabular}{l c c c c}
		\toprule
		Module   & Filters & Filter Size  & Feature Size                          & Number of Nodes    \\
		\midrule
		Input  	& -       & -       &$W\times H\times 3$      & -    \\
		Conv Layer & 64    & $3 \times 64$    & $W\times H\times 64$    & -    \\ 
		Residual Blocks 1-16  & 64    &   $64 \times 64$ & $W\times H\times 64$   & -  \\ 
		GCN Layer    & 64    & $64 \times 64$  & $W\times H\times 64$  &$W*H$ and $L$  \\ 
		Residual Blocks 17-32  & 64    &  $64 \times 64$                & $W\times H\times 64$ & -  \\
		Conv Layer & 3 & $64 \times 3$ &$W\times H\times 3$ & - \\ 
		\bottomrule
	\end{tabular}
	\label{tab:nconfig}
\end{table*}

By default, the proposed Multi-level Attention-guided Graph Network (MAGN) consists of one initial convolutional layer, 16 residual blocks, three graph neural network (GNN) modules, another 16 residual blocks, and one final convolutional layer. The initial and final convolutional layers are used to standardize the input and output data, respectively. The 32 residual blocks are designed to extract local information from the image, while the three GNN modules, connected in series, are employed to capture global information. The detailed configuration is shown in Table~\ref{tab:nconfig}.

When unfolding the feature map, the stride is set to 4 in both vertical and horizontal directions, with a sliding window size of [7, 7]. The output of each convolutional and fully connected layer is passed through the PReLU activation function for non-linear mapping.

The network is trained using the Adam optimizer with a batch size of 32 images, a learning rate of $1e-4$, $\beta$ parameters of $0.9$ and $0.999$, a momentum of $0.9$, and a total of 200 training epochs. For each task, the experimental results are evaluated on commonly used test sets, and the performance is reported using Peak Signal-to-Noise Ratio (PSNR) and Structural Similarity~\cite{wang2004image} (SSIM) metrics.

\subsection{Task}
\subsubsection{Synthetic image denoising}
\begin{table*}[!t]
	\centering
	\small
	\caption[Quantitative Results Comparison for Grayscale Image Denoising]{Quantitative results comparison for grayscale image denoising (the best and second-best results are highlighted in \textbf{bold} and \underline{underlined}, respectively).}
	\resizebox{1.9\columnwidth}{!}{
	\begin{tabular}{l c c c c c c c c c}
		\toprule
		Dataset
		&$\delta$    &BM3D~\cite{dabov2007image}          &DnCNN~\cite{zhang2017beyond}        &FFDNet~\cite{zhang2018ffdnet}         &N3Net~\cite{plotz2018neural}  &NLRN~\cite{liu2018non}             &GCDN~\cite{valsesia2020deep}           &DAGL~\cite{mou2021dynamic}                     &MAGN          \\
		\midrule
		\multirow{3}{*}{Set12}
		&15             &32.37/0.8952 &32.86/0.9031 &32.75/0.9027 &33.03/0.9056    &33.16/0.9070  &33.14/0.9072  &\underline{33.28/0.9100}   &\textbf{33.28/0.9140}\\
		
		&25             &29.96/0.8504 &30.44/0.8622 &30.43/0.8634 &30.55/0.8648 &30.80/0.8689  &30.78/0.8687  &\underline{30.93/0.8720}   &\textbf{30.95/0.8773}   \\
		
		&50             &26.70/0.7676 &27.19/0.7829 &27.31/0.7903 &27.43/0.7948    &27.64/0.7980  &27.60/0.7957  &\underline{27.81/0.8042}   &\textbf{27.82/0.8113}\\
		\midrule
		
		\multirow{3}{*}{BSD68}
		&15             &31.07/0.8717 &31.73/0.8907 &31.63/0.8902 &31.78/0.8927    &31.88/0.8932  &31.83/0.8933  &\underline{31.93/0.8953}   &\textbf{31.94/0.8984}\\
		
		&25             &28.57/0.8013 &29.23/0.8278 &29.19/0.8289 &29.30/0.8321    &29.41/0.8331  &29.35/0.8332  &\underline{29.46/0.8366}   &\textbf{29.52/0.8409}\\
		
		&50             &25.62/0.6864 &26.23/0.7189 &26.29/0.7345 &26.39/0.7293    &26.47/0.7298  &26.38/\underline{0.7389}  &\underline{26.51}/0.7334   &\textbf{26.58/0.7411}\\
		\midrule
		
		\multirow{3}{*}{Urban100}
		&15             &32.35/0.9220 &32.68/0.9255 &32.43/0.9273 &33.08/0.9333    &33.45/0.9354  &33.47/0.9358  &\underline{33.79/0.9393}   &\textbf{33.81/0.9406}\\
		
		&25             &29.71/0.8777 &29.97/0.8797 &29.92/0.8887 &30.19/0.8925    &30.94/0.9018  &30.95/0.9020  &\underline{31.39/0.9093}   &\textbf{31.40/0.9106}\\
		
		&50             &25.95/0.7791 &26.28/0.7874 &26.52/0.8057 &26.82/0.8184    &27.49/0.8279  &27.41/0.8160  &\underline{27.97/0.8423}   &\textbf{27.99/0.8437}\\
		\midrule
		Network Param.
		&                 &-                  &1.56M           &0.49M           &0.72M             &0.35M             &5.99M                &5.62M               &5.22M   \\
		\bottomrule
	\end{tabular}}
	\label{tab:sid}
\end{table*}

\begin{figure}[!t]
	\centering
	\includegraphics[width=0.47\textwidth]{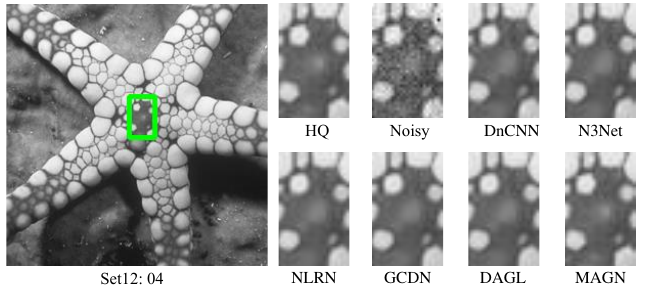}
	\caption[Visual Comparison of Denoising Results]{Visual comparison of denoising results (noise level $\sigma=15$).}
	\label{fig:result1}
\end{figure}

To validate the effectiveness of the proposed Multi-level Attention-guided Graph Network (MAGN), it is compared with several state-of-the-art denoising methods, including BM3D~\cite{dabov2007image}, DnCNN~\cite{zhang2017beyond}, FFDNet~\cite{zhang2018ffdnet}, N3Net~\cite{plotz2018neural}, NLRN~\cite{liu2018non}, GCDN~\cite{valsesia2020deep}, and DAGL~\cite{mou2021dynamic}. Experiments are conducted on the Urban100, BSD68, and Set12 datasets. Following the referenced methods, Gaussian white noise with different levels ($\delta: 15, 25, 50$) is added to clean images to create test sets with varying noise levels, while keeping other experimental parameters consistent. The quantitative results (PSNR and SSIM) and network parameter sizes for all methods under different noise levels are summarized in Table~\ref{tab:sid}. Additionally, visual comparison results are shown in Figure~\ref{fig:result1}.

From the quantitative results in the table, the following observations can be made:
\begin{itemize}
	\item Classical Denoising Methods: BM3D~\cite{dabov2007image}, DnCNN~\cite{zhang2017beyond}, and FFDNet~\cite{zhang2018ffdnet} demonstrate strong performance across all three datasets. For example, DnCNN achieves PSNR values of 32.86, 31.73, and 32.68 on Set12, BSD68, and Urban100, respectively, at a noise level of $\sigma=15$. With only 1.56M parameters, DnCNN exhibits highly efficient representation capabilities.
	
	\item Non-Local Denoising Methods: N3Net~\cite{plotz2018neural} and NLRN~\cite{liu2018non} show significant improvements over classical methods. These methods leverage global information to update and enhance local features during feature fusion and information propagation, thereby improving network performance. For instance, NLRN outperforms DnCNN on the Urban100 dataset by 0.77, 0.97, and 1.21 PSNR at noise levels of 15, 25, and 50, respectively.
	
	\item Graph-Based Methods: GCDN~\cite{valsesia2020deep}, DAGL~\cite{mou2021dynamic}, and MAGN exhibit excellent performance in denoising tasks. Graph structures directly represent non-local relationships in images, enabling flexible aggregation of global information and enhancing model expressiveness. For example, at a noise level of 15, GCDN achieves PSNR/SSIM values of 33.14/0.9072, 31.83/0.8933, and 33.47/0.9358 on Set12, BSD68, and Urban100, respectively, outperforming BM3D, DnCNN, and FFDNet. Similarly, DAGL and MAGN show comparable or superior performance across all datasets. For instance, MAGN achieves 33.28/0.9140, 31.94/0.8984, and 33.81/0.9406 on Set12, BSD68, and Urban100, respectively, at $\sigma=15$.
	
	\item Proposed MAGN Method: MAGN achieves the best performance across all three datasets. Compared to similar graph-based methods like DAGL, MAGN demonstrates further improvements. By effectively integrating local structural information and global representation information, MAGN enhances the network's representation capabilities, leading to superior experimental results.

\end{itemize}

\subsubsection{Image compression artifact reduction}
\begin{table*}[!t]
	\centering
	\small
	\caption[Quantitative Results Comparison for Compression Artifact Reduction]{Quantitative results comparison for compression artifact reduction (the best and second-best results are highlighted in \textbf{bold} and \underline{underlined}, respectively).}
	\resizebox{1.9\columnwidth}{!}{
	\begin{tabular}{l c c c c c c c c c c}
		\toprule
		Dataset 		&$q$   &JPEG            &SA-DCT~\cite{foi2007pointwise}        &ARCNN~\cite{dong2015compression}        &TNRD~\cite{chen2016trainable}           &DnCNN~\cite{zhang2017beyond}       &RNAN~\cite{zhang2019residual}           &DUN~\cite{fu2021model}           &DAGL~\cite{mou2021dynamic}        &MAGN    \\
		\midrule
		
		\multirow{4}{*}{LIVE1}
		&10 &27.77/0.7905 &28.86/0.8093 &28.98/0.8076 &29.15/0.8111 &29.19 /0.8123  &29.63/0.8239  &29.61/0.8232 &\underline{29.70/0.8245}   &\textbf{29.86/0.8252}\\
		
		&20 &30.07/0.8683 &30.81/0.8781 &31.29/0.8733 &31.46/0.8769 &31.59/0.8802   &32.03/0.8877  &31.98/0.8869 &\underline{32.12/0.8887}   &\textbf{32.27/0.8921}\\
		
		&30 &31.41/0.9000 &32.08/0.9078 &32.69/0.9043 &32.84/0.9059 &32.98/0.9090    &33.45/0.9149  &33.38/0.9142 &\underline{33.54/0.9156}   &\textbf{33.68/0.9164}\\
		
		&40 &32.35/0.9173 &32.99/0.9240 &33.63/0.9198 &-/-                &33.96/0.9247    &34.47/0.9299  &34.32/0.9289 &\underline{34.53/0.9305}   &\textbf{34.67/0.9307}\\
		\midrule
		
		\multirow{4}{*}{Classic5}
		&10 &27.82/0.7800 &28.88/0.8071 &29.04/0.7929 &29.28/0.7992 &29.40/0.8026   &29.96/0.8178 &29.95/0.8171 &\underline{30.08/0.8196}   &\textbf{30.10/0.8259} \\
		
		&20 &30.12/0.8541 &30.92/0.8663 &31.16/0.8517 &31.47/0.8576 &31.63/0.8610   &32.11/0.8693 &32.11/0.8689 &\underline{32.35/0.8719}   &\textbf{32.37/0.8889} \\
		
		&30 &31.48/0.8844 &32.14/0.8914 &32.52/0.8806 &32.74/0.8837 &32.91/0.8861    &33.38/0.8924 &33.33/0.8916 &\underline{33.59/0.8942}   &\textbf{33.62/0.9151}\\
		
		&40 &32.43/0.9011 &33.00/0.9055 &33.34/0.8953 &-/-                &33.77/0.9003  &34.27/0.9061 &34.10/0.9045 &\underline{34.41/0.9069}   &\textbf{34.48/0.9303} \\
		\midrule
		Network Parameters   &       &-  &-  &0.12M &- &0.56M  &8.96M &10.5M &5.62M  &5.22M \\
		\bottomrule
	\end{tabular}}
	\label{tab:icar}
\end{table*}

\begin{figure}[!t]
	\centering
	\includegraphics[width=0.47\textwidth]{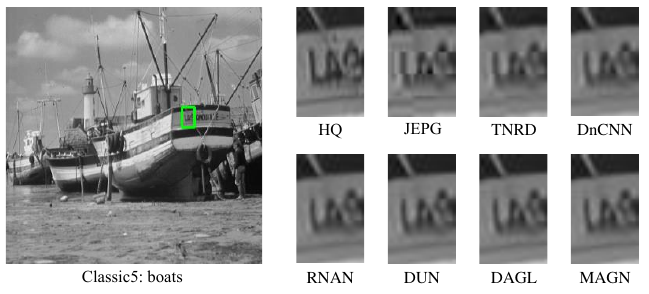}
	\caption[Visual Comparison of Compression Artifact Removal Results]{Visual comparison of compression artifact removal results (quality factor $q=10$).}
	\label{fig:result2}
\end{figure}

To further validate the superiority and effectiveness of the proposed MAGN method, experiments are conducted on the task of compression artifact removal. The proposed method is compared with several classical and advanced approaches, including SA-DCT~\cite{foi2007pointwise}, ARCNN~\cite{dong2015compression}, TNRD~\cite{chen2016trainable}, DnCNN~\cite{zhang2017beyond}, RNAN~\cite{zhang2019residual}, DUN~\cite{fu2021model}, and DAGL~\cite{mou2021dynamic}. The experimental setup follows the referenced methods, ensuring consistent parameter settings. For training and testing, compressed images are generated using the standard Matlab JPEG encoder with quality factors $q: 10, 20, 30, 40$. The quantitative results (PSNR and SSIM) and network parameter sizes for all methods under different compression qualities are summarized in Table~\ref{tab:icar}. Additionally, visual comparison results are shown in Figure~\ref{fig:result2}.

From the quantitative results in the table, the following observations can be made:
\begin{itemize}
	\item Classical Methods: SA-DCT~\cite{foi2007pointwise}, ARCNN~\cite{dong2015compression}, and TNRD~\cite{chen2016trainable} have relatively small network parameters and demonstrate strong performance in compression artifact removal. For example, on the Classic5 dataset, TNRD achieves the best performance with PSNR values of 29.28, 31.47, and 32.74 at quality factors $q=10$, $20$, and $30$, respectively, showing significant improvement over the original JPEG images.
	
	\item Neural Network Methods: DnCNN~\cite{zhang2017beyond}, RNAN~\cite{zhang2019residual}, and DUN~\cite{fu2021model} exhibit superior performance in compression artifact removal, albeit with larger network parameters. For instance, RNAN achieves a PSNR of 29.63 on the LIVE1 dataset at $q=10$, outperforming TNRD by 0.48. Meanwhile, DUN, with 10.5M parameters, demonstrates enhanced representation capabilities due to its larger model size.
	
	\item Graph-Based Methods: DAGL~\cite{mou2021dynamic} and MAGN achieve the best performance. Graph-based methods leverage the advantages of graph structure for information propagation and aggregation, enabling the network to capture non-local image information effectively and improve overall performance.
	
	\item Proposed MAGN Method: MAGN demonstrates the best performance, particularly in terms of SSIM. On the Classic5 dataset, MAGN achieves significant improvements of 0.0063, 0.0170, 0.0209, and 0.0234 in SSIM at quality factors $q=10$, $20$, $30$, and $40$, respectively, compared to the second-best method. This highlights MAGN's ability to enhance network learning and representation by integrating information at two levels, demonstrating its superiority and effectiveness. 
\end{itemize}

\subsubsection{Image demosaicing}
 \begin{table}[!h]
 	\centering
 	\small
 	\caption[Quantitative Results Comparison for Image Demosaicing]{Quantitative results comparison for image demosaicing.}
 	\resizebox{0.9\columnwidth}{!}{
 	\begin{tabular}{l c c c c}
 		\toprule
 		Method &Network Parameters &McMaster18  &Kodak24     &Urban100 \\
 		\midrule
 		Mosaic &-         &9.17/0.1674   &8.56/0.0682  &7.48/0.1195 \\
 		IRCNN~\cite{zhang2017learning} &0.19M &37.47/0.9615 &40.41/0.9807 &36.64/0.9743 \\
 		RNAN~\cite{zhang2019residual}  &8.96M &39.71/0.9725 &43.09/0.9902 &39.75/0.9848 \\
 		DAGL~\cite{mou2021dynamic}   &5.62M &\underline{39.84/0.9735} &\underline{43.21/0.9910} &\underline{40.20/0.9854} \\
 		\midrule
 		MAGN &5.22M    &\textbf{40.03/0.9745}  &\textbf{43.29/0.9910} &\textbf{40.30/0.9857} \\
 		\bottomrule
 	\end{tabular}}
 	\label{tab:id}
 \end{table}
 
 \begin{figure}[!h]
 	\centering
 	\includegraphics[width=0.45\textwidth]{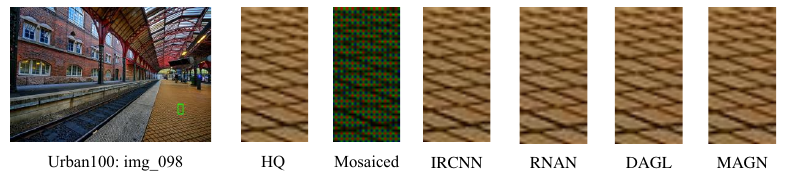}
 	\caption[Visual Comparison of Demosaicing Results]{Visual comparison of demosaicing results.}
 	\label{fig:result3}
 \end{figure}
 
 To further validate the effectiveness of the proposed method, experiments are conducted on the image demosaicing task. In this task, the proposed MAGN method is compared with RNAN~\cite{zhang2019residual}, IRCNN~\cite{zhang2017learning}, and DAGL~\cite{mou2021dynamic}. The experimental setup follows the referenced methods, ensuring consistent parameter settings. For training and testing, the same mosaicking process is applied to the images, and other experimental configurations remain unchanged. The quantitative results (PSNR and SSIM) and network parameter sizes for all methods are summarized in Table~\ref{tab:id}. Additionally, visual comparison results are shown in Figure~\ref{fig:result3}.
 
 From the quantitative results in the table, the following observations can be made:
 \begin{itemize}
 	\item Neural Network Methods: RNAN~\cite{zhang2019residual} and IRCNN~\cite{zhang2017learning} demonstrate strong performance in image demosaicing. Both methods achieve PSNR values exceeding 30 across all three datasets, indicating that the processed images effectively retain the original image information. Notably, on the Kodak24 dataset, RNAN achieves a PSNR of 43.09, showcasing its ability to effectively remove mosaicking artifacts.
 	
 	\item Graph-Based Methods: DAGL~\cite{mou2021dynamic} and MAGN exhibit superior performance in image demosaicing. Compared to classical neural network methods, graph-based methods leverage the advantages of graph structure for information updating and aggregation, leading to enhanced performance.

 	\item Proposed MAGN Method: MAGN achieves the best results, with PSNR values of 40.03, 43.29, and 40.30 on the McMaster18, Kodak24, and Urban100 datasets, respectively. These results consistently exceed 40, demonstrating MAGN's robust performance. With a manageable parameter size of 5.22M, MAGN effectively integrates global and local representations and structures, enhancing the network's representation capabilities and delivering superior experimental outcomes.
 \end{itemize}

\subsection{Ablation study}
To better demonstrate the relationship between the performance of the MAGN method and its parameters, experiments were conducted on the image demosaicing task to validate the following aspects:  
i) the impact of different network structures on performance, and  
ii) the influence of different local structure scales on performance.  

In the MAGN method, two types of graphs are constructed: a global representation graph and a local structure graph. The information propagation and aggregation in these graphs significantly affect the network's performance. Additionally, when constructing node information in the local structure graph, a fixed-scale sliding window is used. Different window sizes lead to variations in graph information, which in turn affect the network's performance. The following sections discuss the impact of these parameters on network performance and provide a quantitative analysis through comparative experiments.

\subsubsection{Effectiveness Analysis of Network Structures}
\begin{table}[!h]
	\centering
	\small
	\caption[Quantitative Results Comparison of Different Network Structures on Image Demosaicing]{Quantitative results comparison of different network structures on image demosaicing.}
	\resizebox{1\columnwidth}{!}{
	\begin{tabular}{l c c c c c}
		\toprule
		Method      &Global Graph  &Local Graph  &McMaster18    &Kodak24           &Urban100 \\
		\midrule
		w/o Global    & &$\checkmark$ &39.91/0.9740   &43.19/0.9909   &40.14/0.9854   \\
		w/o Local  &$\checkmark$ & &39.49/0.9720   &42.83/0.9903   &39.46/0.9841   \\
		w/ Multiple Local & &$\checkmark$ $\checkmark$ &39.79/0.9733 &43.16/0.9908  &39.99/0.9850  \\
		MAGN   &$\checkmark$ &$\checkmark$ &\textbf{40.03/0.9745}  &\textbf{43.29/0.9910} &\textbf{40.30/0.9857} \\
		\bottomrule
	\end{tabular}}
	\label{tab:framework}
\end{table}

The experiments were conducted by training different network structures on the DIV2K dataset and testing them on the McMaster18, Kodak24, and Urban100 datasets for image demosaicing. Four network structures were designed:  
i) a network with only the local structure graph (w/o Global),  
ii) a network with only the global representation graph (w/o Local),  
iii) a network with two local structure graphs (w/ Multiple Local), and  
iv) the proposed MAGN network, which integrates both global and local graphs.  

The quantitative results (PSNR and SSIM) for different network structures are summarized in Table~\ref{tab:framework}.  

From the data in Table~\ref{tab:framework}, it can be observed that both global and local information contribute to the performance of image demosaicing. While global representation information enhances the network's feature extraction capabilities, its impact is less pronounced compared to local structure information. Additionally, using multiple local graphs does not necessarily improve performance, as redundant or conflicting information may degrade results. The MAGN method, which combines global and local information, achieves the best performance across all three datasets, demonstrating the complementary benefits of integrating both types of information. The flexible graph structure enables the network to capture hierarchical information, enhancing its representation capabilities and delivering superior results.

\subsubsection{Effectiveness Analysis of Local Structure Scales}
\begin{table}[!h]
	\centering
	\small
	\caption[Quantitative Results Comparison of Different Local Structure Scales on Image Demosaicing]{Quantitative results comparison of different local structure scales on image demosaicing.}
	\resizebox{0.9\columnwidth}{!}{
	\begin{tabular}{l c c c c }
		\toprule
		Method      &Local Scale  &McMaster18    &Kodak24           &Urban100 \\
		\midrule
		MAGN    &5  &39.79/0.9733   &43.04/0.9901   &39.79/0.9846   \\
		MAGN    &7  &\textbf{40.03/0.9745}  &\textbf{43.29/0.9910} &\textbf{40.30/0.9857} \\
		MAGN    &9  &39.90/0.9738   &43.17/0.9907   &40.10/0.9852  \\
		MAGN    &11  &39.78/0.9732   &43.07/0.9903   &39.97/0.9849   \\
		\bottomrule
	\end{tabular}}
	\label{tab:kernelsize}
\end{table}

Four different sliding window sizes ($5, 7, 9, 11$) were used to construct local structure graphs, while keeping other configurations consistent. The networks were trained on the DIV2K dataset and tested on the McMaster18, Kodak24, and Urban100 datasets. The quantitative results (PSNR and SSIM) for different local structure scales are summarized in Table~\ref{tab:kernelsize}.  

From the data in Table~\ref{tab:kernelsize}, it can be observed that network performance does not consistently improve with increasing window size. Instead, the best performance is achieved with a window size of 7. Larger window sizes may introduce redundant information and noise, while smaller sizes may fail to capture sufficient structural information. Thus, a window size of 7 strikes a balance between learning complexity and effective feature transmission, enabling the network to achieve optimal performance.

\subsubsection{Network Parameters and Attention Mechanism Analysis}
The MAGN network has 5.22M parameters, primarily from the residual and graph network modules. Despite constructing two graphs and integrating global and local information, MAGN has fewer parameters than the DAGL method~\cite{mou2021dynamic}. The streamlined and optimized network structure enhances its efficiency and representation capabilities. Additionally, MAGN employs a multi-head attention mechanism to guide the learning of graph structure relationships, enabling diverse node relationship matrices and improving information propagation and representation. The network is trained using the Adam optimizer, with stable loss convergence and no issues of gradient explosion or vanishing. Performance improves steadily with training epochs, and parameters converge quickly.

\subsubsection{Network Learning Rate Analysis}
\begin{figure}[!h]
	\centering
	\includegraphics[width=0.4\textwidth]{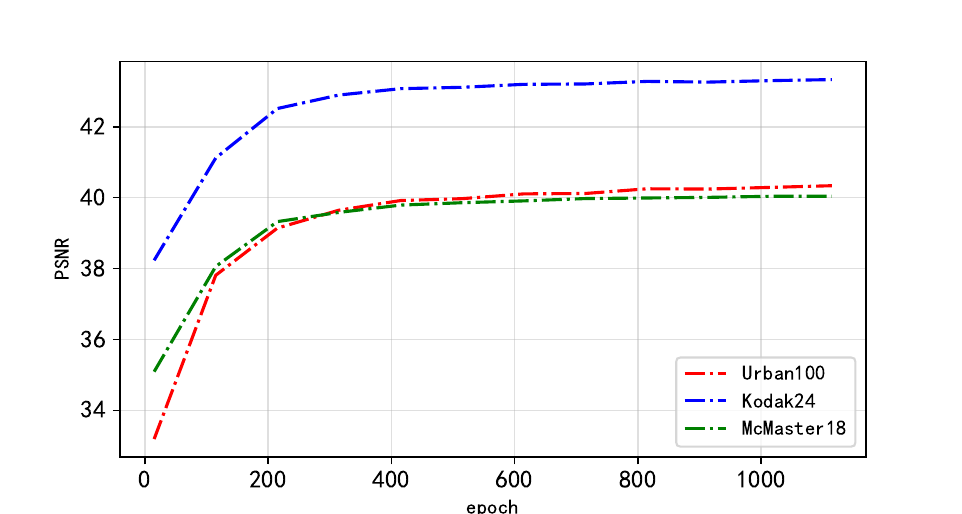}
	\caption[Performance Change of MAGN Across Training Epochs (PSNR)]{Performance change of MAGN across training epochs (PSNR).}
	\label{fig:speed1}
\end{figure}
\begin{figure}[!h]
	\centering
	\includegraphics[width=0.4\textwidth]{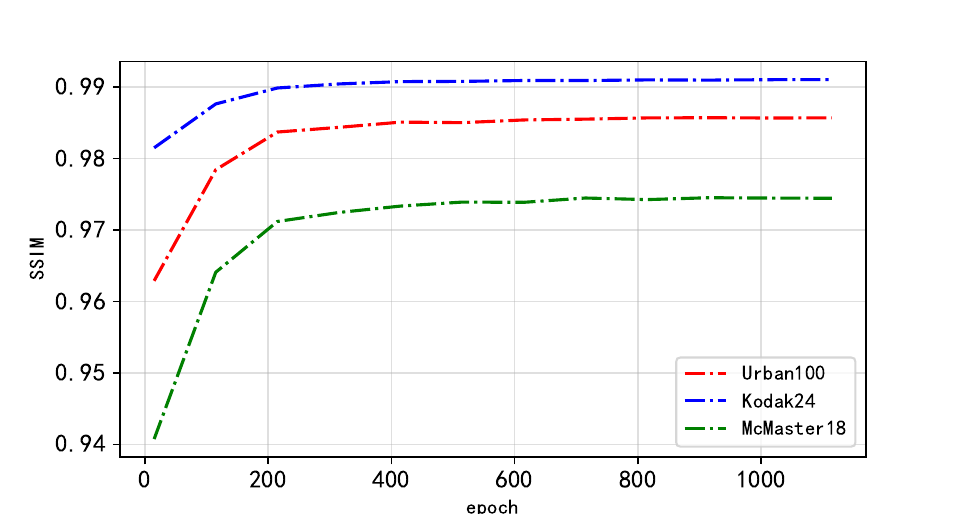}
	\caption[Performance Change of MAGN Across Training Epochs (SSIM)]{Performance change of MAGN across training epochs (SSIM).}
	\label{fig:speed2}
\end{figure}
\begin{figure}[!t]
	\centering
	\includegraphics[width=0.45\textwidth]{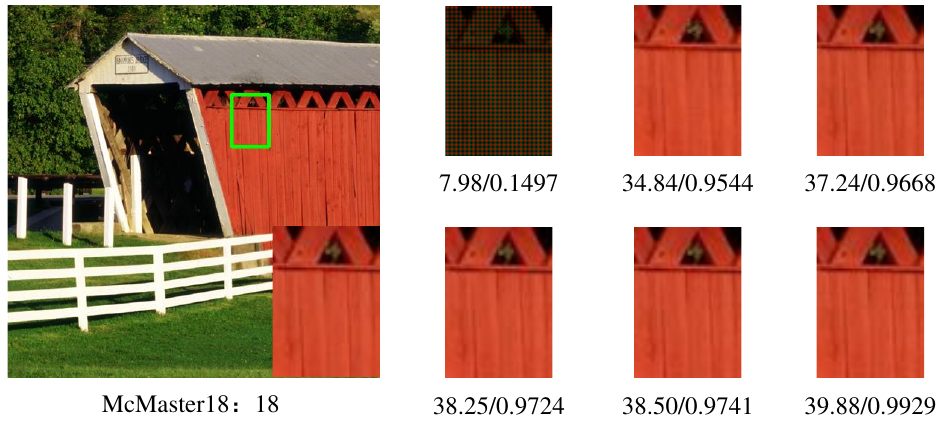}
	\caption[Results on McMaster18 Dataset with Different Network Performances]{Results on McMaster18 dataset with different network performances.}
	\label{fig:example1}
\end{figure}
\begin{figure}[!t]
	\centering
	\includegraphics[width=0.45\textwidth]{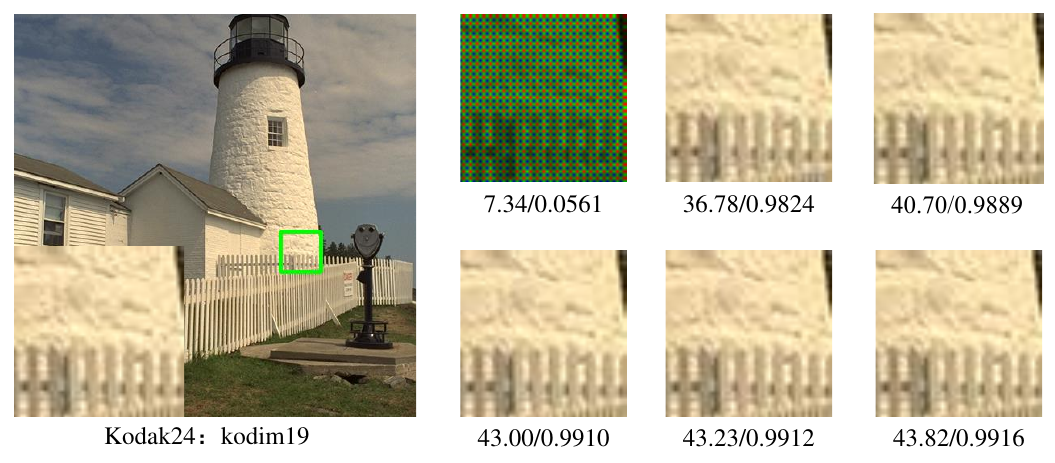}
	\caption[Results on Kodak24 Dataset with Different Network Performances]{Results on Kodak24 dataset with different network performances.}
	\label{fig:example2}
\end{figure}
\begin{figure}[!t]
	\centering
	\includegraphics[width=0.45\textwidth]{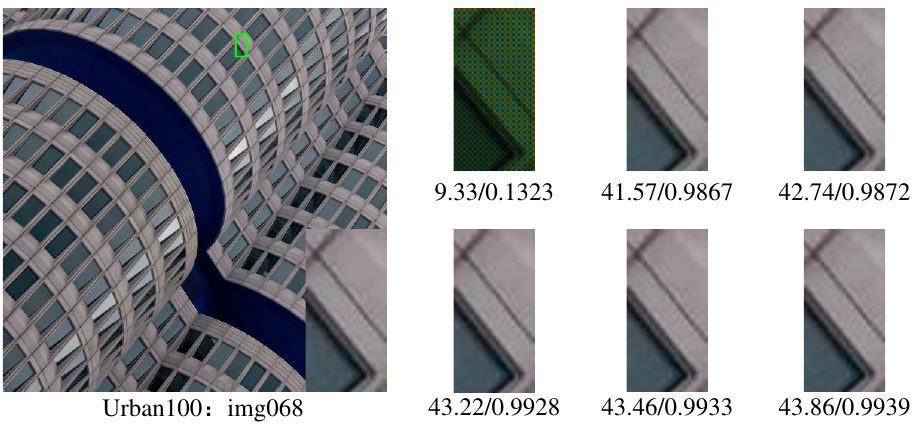}
	\caption[Results on Urban100 Dataset with Different Network Performances]{Results on Urban100 dataset with different network performances.}
	\label{fig:example3}
\end{figure}

The learning speed of MAGN is comparable to that of the reference method~\cite{mou2021dynamic}, primarily determined by the learning speed of the multi-layer residual and graph structure modules. Figures~\ref{fig:speed1} and~\ref{fig:speed2} show the performance change curves across training epochs. As training progresses, performance initially improves rapidly, then slows, and eventually reaches a plateau. Example results on the McMaster18, Kodak24, and Urban100 datasets are shown in Figures~\ref{fig:example1},~\ref{fig:example2}, and~\ref{fig:example3}. As network performance improves, missing pixels in the mosaicked images are gradually restored to values closer to the original images, resulting in more realistic outputs.

\section{Conclusion}
In this paper, we proposes an attention-guided graph convolutional neural network designed to integrate global representation information and local structural information in images, and applies it to image restoration tasks. Specifically, the method employs an attention mechanism to dynamically construct and optimize the graph structure within the network during training. By performing graph convolution operations on the constructed graph structure, the method extracts global representation information from images, addressing the limitations of traditional convolutional networks in capturing global context. Furthermore, the attention-guided graph convolutional structure can be seamlessly embedded into any convolutional network, enabling end-to-end training. Unlike previous graph convolutional neural networks, the proposed architecture considers global information at two levels within the feature map, allowing the network to more comprehensively integrate global representation and local structural information. Experimental results on image restoration tasks demonstrate the effectiveness and superiority of the proposed graph convolutional neural network.

\ifCLASSOPTIONcaptionsoff
  \newpage
\fi

\begin{small}
  \bibliographystyle{IEEETran}
  \bibliography{myrefs}
\end{small}
\vskip -0.8in

\end{document}